\documentclass[twocolumn,superscriptaddress,pra]{revtex4}
\usepackage{amsmath}
\usepackage{amssymb}
\usepackage{graphicx}
\usepackage{bm}

\begin{document}

\title{Quantum Fluctuations of the Stress-Energy Tensor via Noether's Theorem and the Connection to the Cosmological Constant}

\author{J. A. Crosse}
\email{jac32@nyu.edu}
\affiliation{New York University Shanghai, 567 West Yangsi Road, Shanghai, 200124, China.}
\affiliation{NYU-ECNU Institute of Physics at NYU Shanghai, 3663 Zhongshan Road North, Shanghai, 200062, China.}

\date{\today}

\begin{abstract}
When deriving the expression for the conserved Noether's current one obtains a term that is an integral over the Euler-Lagrange equation. This term vanishes vanishes if the system evolves along the classical path. However, quantum particles do not necessarily follow the classical path and, hence, for a quantum system, the Euler-Lagrange term does not vanish. This extra term corresponds to a quantum fluctuation of the Noether's current owing to the system deviating from the classical path. Here, we compute the fluctuation of a quantum field directly from Noether's Theorem and, by considering a spacetime translational symmetry of a complex scalar field, we find the specific fluctuations of the Stress-Energy tensor. As the Stress-Energy tensor is the source term for Einstein's equations of General Relativity, these fluctuations can be seen to be a contribution to the Cosmological Constant. The final expression for this contribution is proportional to the mass of the field to the fourth power which is consistent with previously derived expressions for the Cosmological Constant.
\end{abstract} 

\maketitle

\section{Introduction}

It has been shown that by relaxing the constraint that a physical system must follow the classical path, as defined by the solutions to the Euler-Lagrange Equation, it is possible to obtain an extra term in the Noether's current \cite{VF1}. This term corresponds to fluctuations in the current owing to the system deviating from the classical path. Using Feynman's path integral formalism \cite{path}, this extra term can be evaluated. For time translation symmetry in the classical action of Harmonic oscillator, it is possible  to recover the energy features of the quantum Harmonic oscillator, including the energy-time commutation relations and the zero point energy \cite{VF1}. 

Here we extend this result to field theory \cite{sup}. By relaxing the constraint that the system must follow the classical path and evaluating the weighted contribution of all possible paths, we recover an extra term in the Stress-Energy tensor that corresponds to a quantum fluctuation. As the Stress-Energy tensor is the source term for Einstein's equations of General Relativity, the extra term can be associated with the Cosmological Constant.

\section{Noether's Theorem and the Stress-Energy Tensor}

Noether's Theorem \cite{noether} states that if the action, $S$, is invariant under some transformation then there is an associated conserved quantity. In classical field theory, if we vary the field
\begin{equation}
\phi \rightarrow \phi' = \phi + \alpha\delta\phi,
\label{phivary}
\end{equation}
and the action, $S$, or equivalently the Lagrangian density, $\mathcal{L}$, is unchanged up to a divergence (surface term)
\begin{equation}
\mathcal{L}(\phi) \rightarrow \mathcal{L}'(\phi') = \mathcal{L}(\phi') - \alpha\partial_{\mu}J^{\mu},
\label{Lvary}
\end{equation}
then we can find some combinations of parameters that remains unchanged as the field evolves. We derive an expression for this conserved quantity by varying action
\begin{equation}
\delta S = \int d^{4}x\,\left\{\frac{\partial \mathcal{L}}{\partial \phi}\alpha\delta\phi + \frac{\partial \mathcal{L}}{\partial (\partial_{\mu}\phi)}\delta(\alpha\partial_{\mu}\phi)\right\}= \alpha\partial_{\mu}J^{\mu}.
\end{equation}
After some rearrangement one finds
\begin{multline}
\alpha\int d^{4}x\,\partial_{\mu}\left\{\left(\frac{\partial \mathcal{L}}{\partial (\partial_{\mu}\phi)}\delta\phi\right)-J^{\mu}\right\}\\
+ \alpha\int d^{4}x\,\left\{\frac{\partial \mathcal{L}}{\partial \phi} - \partial_{\mu}\left(\frac{\partial \mathcal{L}}{\partial (\partial_{\mu}\phi)}\right)\right\}\delta\phi = 0.
\label{noether}
\end{multline}
The second term vanishes because the integrand is just the Euler-Lagrange equation which is identically zero along the classical path
\begin{equation}
\frac{\partial \mathcal{L}}{\partial \phi} - \partial_{\mu}\left(\frac{\partial \mathcal{L}}{\partial (\partial_{\mu}\phi)}\right) = 0.
\end{equation}
For the action, $S$, to be unchanged by the variation the first term must also vanish
\begin{equation}
\alpha\int d^{4}x\,\partial_{\mu}\left\{\left(\frac{\partial \mathcal{L}}{\partial (\partial_{\mu}\phi)}\delta\phi\right)-J^{\mu}\right\} = 0.
\end{equation}
Thus, the integrand must vanish
\begin{equation}
\partial_{\mu}\left\{\left(\frac{\partial \mathcal{L}}{\partial (\partial_{\mu}\phi)}\delta\phi\right)-J^{\mu}\right\} = 0,
\end{equation}
and the expression within the derivative must be a constant of motion
\begin{equation}
j^{\mu} = \left(\frac{\partial \mathcal{L}}{\partial (\partial_{\mu}\phi)}\delta\phi\right)-J^{\mu}.
\end{equation}
The quantity, $ j^{\mu}$, is known as the Noether's current.

If one computes the conserved quantity when the action, $S$, is unchanged by a spacetime transformation
\begin{equation}
x^{\mu} \rightarrow x'^{\mu} = x^{\mu} + a^{\mu},
\end{equation}
one arrives at the Stress-Energy tensor
\begin{equation}
T^{\mu}_{\;\;\nu} = \left(\frac{\partial \mathcal{L}}{\partial (\partial_{\mu}\phi)}\right)\partial_{\nu}\phi-\eta_{\;\;\nu}^{\mu}\mathcal{L},
\end{equation}
where $\eta_{\;\;\nu}^{\mu}$ is the flat space Minkowski metric.

\section{Quantum Fluctuations of the Stress-Energy Tensor}

The above analysis assumes that the system follows the classical path. However, Feynman's path integral quantization formalism implies that quantum fields do not necessarily evolve this way \cite{path}. In fact, the systems evolution is the weighted average over all possible paths. For these non-classical paths the evolution of the field would not be described by the Euler-Lagrange equation and, hence, the second term in Eq. \eqref{noether} would not vanish 
\begin{equation}
\alpha\int d^{4}x\,\partial_{\mu}F^{\mu} = \alpha\int d^{4}x\,\left\{\frac{\partial \mathcal{L}}{\partial \phi} - \partial_{\mu}\left(\frac{\partial \mathcal{L}}{\partial (\partial_{\mu}\phi)}\right)\right\}\delta\phi \neq 0.
\label{f1}
\end{equation}
In this case $F^{\mu}$ would also contribute to the conserved current and will be a fluctuation in the Noether's current owing to the field not evolving along the classical path
\begin{equation}
j^{\mu} = \left(\frac{\partial \mathcal{L}}{\partial (\partial_{\mu}\phi)}\delta\phi\right)-J^{\mu} + F^{\mu}.
\label{j}
\end{equation}

In order to compute $F^{\mu}$ one can functionally expand the action, $S$, about the field values that solve the Euler-Lagrange Equation (i.e. the classical path)
\begin{widetext}
\begin{align}
S(\phi_{cl}+\Delta\phi) &= S(\phi_{cl}) + \int d^{4}x' \,\Delta\phi(x')\left.\frac{\delta S}{\delta \phi(x')}\right|_{\phi_{cl}} + \frac{1}{2!}\int d^{4}x'\int d^{4}x''\, \Delta\phi(x')\Delta\phi(x'')\left.\frac{\delta^{2} S}{\delta \phi(x')\delta \phi(x'')}\right|_{\phi_{cl}} + \cdots.
\end{align}
\end{widetext}
The zeroth order term is just the action along the classical path, which obeys the Euler-Lagrange Equation and so it's contribution to the fluctuations, $F^{\mu}$, vanishes. The first order term is the variation of the action which, by definition, is equal to zero along the classical path. Hence, the lowest order term to contribute to the fluctuations is the second order term. If the Lagrangian density is bilinear or quadratic in the field and its derivative, the second order functional derivatives will result in two times the constant prefactor for that term, which cancels the factor of half from the functional expansion. Hence, the effect of the expansion is to replace the field, $\phi(x)$, in the Lagrangian density with the deviation of the field from the classical value, $\Delta\phi(x)$,
\begin{equation}
\mathcal{L}\left[\phi(x),\partial_{\mu}\phi(x)\right] \rightarrow \mathcal{L}\left[\Delta\phi(x),\partial_{\mu}\Delta\phi(x)\right].
\label{L2}
\end{equation}
Furthermore, this term will be the highest order term in the expansion. All higher order terms will vanish. 

One can compute the fluctuations, $F^{\mu}$, by substituting Eq. \eqref{L2} into Eq. \eqref{f1} and changing variables $\phi(x) \rightarrow \phi_{cl}(x) + \Delta\phi(x)$ where $\phi_{cl}(x)$ is the classical value of the field and acts as a constant. Thus
\begin{multline}
\alpha\int d^{4}x\,\partial_{\mu}F^{\mu} =\\
\alpha\int d^{4}x\,\left\{\frac{\partial \mathcal{L}\left[\Delta\phi\right]}{\partial \Delta\phi} - \partial_{\mu}\left(\frac{\partial \mathcal{L}\left[\Delta\phi\right]}{\partial (\partial_{\mu}\Delta\phi)}\right)\right\}\delta\Delta\phi.
\end{multline}
One can simplify this by noting that 
\begin{equation}
\partial_{\nu}\mathcal{L} = \frac{\partial\mathcal{L}}{\partial\Delta\phi}\partial_{\nu}\Delta\phi + \frac{\partial\mathcal{L}}{\partial(\partial_{\mu}\Delta\phi)}\partial_{\nu}\left(\partial_{\mu}\Delta\phi\right),
\end{equation}
and recalling that for a spacetime translation $\alpha \delta \Delta\phi =  a^{\nu}\partial_{\nu}\Delta\phi$
\begin{multline}
\alpha\int d^{4}x\,\partial_{\mu}F^{\mu} =\\
-a^{\nu}\partial_{\mu}\int d^{4}x\,\left\{\left(\frac{\partial\mathcal{L}\left[\Delta\phi\right]}{\partial(\partial_{\mu}\Delta\phi)}\partial_{\nu}\Delta\phi\right)-\eta_{\;\;\nu}^{\mu}\mathcal{L}\left[\Delta\phi\right]\right\}.
\label{f3}
\end{multline}
The expression in the brackets is the Stress-Energy tensor except it is evaluated at the deviation of the field from the classical value, $\Delta\phi(x)$, rather than at the classical field value, $\phi(x)$. Comparing the expression in Eq. \eqref{f3} with Eqs. \eqref{f1} and \eqref{j} one can see the fluctuation, $F_{\;\;\nu}^{\mu}$, is given by
\begin{equation}
F_{\;\;\nu}^{\mu} = -T^{\mu}_{\;\;\nu}(\Delta\phi),
\end{equation} 
which is minus the Stress-Energy tensor evaluated at the distance from the classical path.

\section{Quantum Fluctuations of a Complex Scalar Field}

\subsection{General Expression}

As an example we can compute the Stress-Energy Tensor fluctuations of a free complex scalar field. The Lagrangian density, $\mathcal{L}$, for a complex scalar field is given by
\begin{equation}
\mathcal{L}(\phi,\phi^\ast) = \partial_{\mu}\phi^\ast(x)\partial^{\mu}\phi(x) - m^2\phi^\ast(x)\phi(x),
\end{equation}
where the mass $m$ is in units of $L^{-1}$. The conversion to mass in SI units, $\bar{m}$, is given by $m = \bar{m}c/\hbar$. Expanding the action about the classical field values, $\phi_{cl}$ and $\phi^\ast_{cl}$, gives
\begin{widetext}
\begin{align}
S(\phi_{cl}+\Delta\phi, \phi^\ast_{cl}+\Delta\phi^\ast) = S(\phi_{cl}, \phi_{cl})  &+ \int d^{4}x' \,\Delta\phi^\ast(x')\left.\frac{\delta S}{\delta \phi^\ast(x')}\right|_{\phi^\ast_{cl}, \phi_{cl}} + \int d^{4}x'' \,\Delta\phi(x'')\left.\frac{\delta S}{\delta \phi(x'')}\right|_{\phi^\ast_{cl}, \phi_{cl}}\nonumber\\
&\hspace{2cm} + \int d^{4}x'\int d^{4}x''\, \Delta\phi^\ast(x')\Delta\phi(x'')\left.\frac{\delta^{2} S}{\delta \phi^\ast(x')\delta \phi(x'')}\right|_{\phi^\ast_{cl}, \phi_{cl}}.
\end{align}
Note that, as the action, $S$, is bilinear in the field, $\phi$, and the conjugate field, $\phi^\ast$, so this expansion is exact. The functional derivatives of the action are 
\begin{align}
\left.\frac{\delta S}{\delta \phi^\ast(x')}\right|_{\phi_{cl}, \phi^\ast_{cl}} &= \int d^{4}x\,\left.\left\{\partial_{\mu}\frac{\delta\phi^\ast(x)}{\delta \phi^\ast(x')}\partial^{\mu}\phi(x) - m^2\frac{\delta\phi^\ast(x)}{\delta\phi^\ast(x')}\phi(x)\right\}\right|_{\phi_{cl}, \phi^\ast_{cl}},\nonumber\\
&= \int d^{4}x\,\left.\left\{-\partial_{\mu}\partial^{\mu}\phi(x) - m^2\phi(x)\right\}\delta(x-x')\right|_{\phi_{cl}, \phi^\ast_{cl}}=0,
\end{align}
where in the last line we have used the fact that the expression in the brackets is the Klein-Gordon equation for the field, which vanishes for the classical field. This is expected as the first functional derivative vanishes at the stationary points of the action, which is the definition of the classical path. Similarly
\begin{align}
\left.\frac{\delta S}{\delta \phi(x'')}\right|_{\phi_{cl}, \phi^\ast_{cl}} &= \int d^{4}x\,\left.\left\{\partial_{\mu}\phi^\ast(x)\partial^{\mu}\frac{\delta\phi(x)}{\delta \phi(x'')} - m^2\phi^\ast(x)\frac{\delta\phi(x)}{\delta\phi(x'')}\right\}\right|_{\phi_{cl}, \phi^\ast_{cl}},\nonumber\\\
&= \int d^{4}x\,\left.\left\{-\partial_{\mu}\partial^{\mu}\phi^\ast(x) - m^2\phi^\ast(x)\right\}\delta(x-x'')\right|_{\phi_{cl}, \phi^\ast_{cl}}=0.
\end{align}
Again, the expression in the brackets is the Klein-Gordon equation for the conjugate field, which also vanishes for the classical field and, hence, this contribution also vanishes as expected. Finally
\begin{align}
\left.\frac{\delta^{2} S}{\delta \phi^\ast(x')\delta \phi(x'')}\right|_{\phi_{cl}, \phi_{cl}} &= \int d^{4}x\,\left.\left\{\partial_{\mu}\frac{\delta\phi^\ast(x)}{\delta \phi^\ast(x')}\partial^{\mu}\frac{\delta\phi(x)}{\delta \phi(x'')} - m^2\frac{\delta\phi^\ast(x)}{\delta\phi^\ast(x')}\frac{\delta\phi(x)}{\delta\phi(x'')}\right\}\right|_{\phi_{cl}, \phi^\ast_{cl}},\nonumber\\
&= \int d^{4}x\,\left.\left\{-\partial_{\mu}\partial^{\mu}\delta(x-x'') - m^2\delta(x-x'')\right\}\delta(x-x')\right|_{\phi_{cl}, \phi^\ast_{cl}}.
\end{align}
\end{widetext}
This expression does not vanish and, hence, contributes to the fluctuation in the Noether's current. The fluctuation can be found by computing the second order term in the expansion of the action
\begin{multline}
\int d^{4}x'\int d^{4}x'' \,\Delta\phi^\ast(x')\Delta\phi(x'')\left.\frac{\delta^{2} S}{\delta \phi^\ast(x')\delta \phi(x'')}\right|_{\phi_{cl}, \phi^\ast_{cl}} \\
= \int d^{4}x \,\left\{\partial_{\mu}\Delta\phi^\ast(x)\partial^{\mu}\Delta\phi(x) - m^2\Delta\phi^\ast(x)\Delta\phi(x)\right\}.
\end{multline}

The fluctuation, $F^{\mu}$, in a field $\phi$ is given by
\begin{equation}
\alpha\int d^{4}x\,\partial_{\mu}F^{\mu} = \alpha\int d^{4}x\,\left\{\frac{\partial \mathcal{L}_{2}}{\partial \phi} - \partial_{\mu}\left(\frac{\partial \mathcal{L}_{2}}{\partial (\partial_{\mu}\phi)}\right)\right\}\delta\phi.
\end{equation}
where
\begin{equation}
\mathcal{L}_{2} = \bigg\{\partial_{\mu}\Delta\phi^\ast(x)\partial^{\mu}\Delta\phi(x) - m^2\Delta\phi^\ast(x)\Delta\phi(x)\bigg\}.
\end{equation}
For a spacetime translation one has
\begin{equation}
\alpha = a^{\nu} \qquad\mathrm{and}\qquad \delta\phi = \partial_{\nu}\phi.
\end{equation}
Note that, as we have expanded the field about the classical path, $\phi = \phi_{cl} + \Delta\phi$, which is fixed. The change in the field is given only given by $\Delta\phi$. Hence, 
\begin{equation}
\partial_{\nu}\phi = \partial_{\nu}\Delta\phi \qquad\mathrm{and}\qquad \frac{\partial}{\partial\phi} = \frac{\partial}{\partial\Delta\phi}.
\end{equation}
In this case, there are two fluctuating fields, the field, $\phi$, and the conjugate field, $\phi^\ast$,
\begin{equation}
\alpha\int d^{4}x\,\partial_{\mu}F^{\mu} = f_{\phi} + f_{\phi^\ast},
\label{ft}
\end{equation}
with each contribution having the form
\begin{equation}
f_{\phi^{(\ast)}} = a^{\nu}\int d^{4}x\,\left\{\frac{\partial \mathcal{L}_{2}}{\partial \Delta\phi^{(\ast)}} - \partial_{\lambda}\left(\frac{\partial \mathcal{L}_{2}}{\partial (\partial_{\lambda}\Delta\phi^{(\ast)})}\right)\right\}\partial_{\nu}\Delta\phi^{(\ast)} .
\end{equation}
Evaluating $f_{\phi^{(\ast)}}$ for the two fields gives
\begin{equation}
f_{\phi} = a^{\nu}\int d^{4}x\,\bigg\{-m^2\Delta\phi^\ast\left(\partial_{\nu}\Delta\phi\right) - \left(\partial^{\lambda}\partial_{\lambda}\Delta\phi^\ast\right)\partial_{\nu}\Delta\phi \bigg\},
\end{equation}
and
\begin{equation}
f_{\phi^\ast} = a^{\nu}\int d^{4}x\,\bigg\{-m^2\left(\partial_{\nu}\Delta\phi^\ast\right)\Delta\phi - \left(\partial^{\lambda}\partial_{\lambda}\Delta\phi\right)\partial_{\nu}\Delta\phi^\ast \bigg\}.
\end{equation}
Thus, the full fluctuation is
\begin{align}
&f_{\phi} + f_{\phi^\ast}\nonumber\\
&= a^{\nu}\int d^{4}x\,\bigg\{-\bigg(\left(\partial^{\lambda}\partial_{\lambda}\Delta\phi\right)\partial_{\nu}\Delta\phi^\ast+\left(\partial^{\lambda}\partial_{\lambda}\Delta\phi^\ast\right)\partial_{\nu}\Delta\phi\bigg)\nonumber\\
&\hspace{2.8cm} -m^2\bigg(\Delta\phi^\ast\left(\partial_{\nu}\Delta\phi\right)+\left(\partial_{\nu}\Delta\phi^\ast\right)\Delta\phi\bigg)\bigg\},\nonumber\\
&= a^{\nu}\int d^{4}x\,\bigg\{-\partial_{\mu}\left[\partial^{\mu}\Delta\phi^\ast\partial_{\nu}\Delta\phi + \partial_{\nu}\Delta\phi^\ast\partial^{\mu}\Delta\phi\right.\nonumber\\
&\hspace{2.8cm} \left. - \eta^{\mu}_{\;\;\nu}\left( \partial_{\lambda}\Delta\phi^\ast\partial^{\lambda}\Delta\phi - m^2\Delta\phi^\ast\Delta\phi\right) \right]\bigg\},\nonumber\\
&= a^{\nu}\int d^{4}x\,\bigg\{-\partial_{\mu}\Delta T^{\mu}_{\;\;\nu}(\Delta\phi)\bigg\}.
\label{Dfluc}
\end{align}
Hence, the fluctuation, $F^{\mu}_{\;\;\nu}$, in the Stress-Energy tensor for a complex scalar field is given by
\begin{equation}
F^{\mu}_{\;\;\nu} = -\Delta T^{\mu}_{\;\;\nu}(\Delta\phi),
\end{equation}
\newline 
as expected. Note that the expression for the Stress-Energy Tensor, $\Delta T^{\mu}_{\;\;\nu}(\Delta\phi)$, is necessarily of the canonical form as any symmetrization that relies on the equations of motion for the classical field would not be valid here as the deviations from the classical field, by definition, do not obey the classical equations of motion.

\subsection{Time Averaged Fluctuations}

One can find the average fluctuation in the stress energy tensor in the vacuum, $\Delta T^{\mu}_{\;\;\nu}$, using the path integral formulism. To compute this quantity one can find the matrix element of the energy fluctuations within a spacetime volume and then average over that volume
\begin{align}
\Delta T^{\mu}_{\;\;\nu} &= \frac{1}{L^4}\int d^4x\, \langle -F^{\mu}_{\;\;\nu} \rangle \nonumber\\
&= \frac{1}{L^4}\int d^4x\,\frac{\displaystyle\int\mathcal{D}\Delta\phi^\ast\int\mathcal{D}\Delta\phi\,\left\{\Delta T^{\mu}_{\;\;\nu}(\Delta\phi) \right\} e^{iS/\hbar}}{\displaystyle\int\mathcal{D}\Delta\phi^\ast\int\mathcal{D}\Delta\phi \,e^{iS/\hbar}}.
\end{align}
We can evaluate the path integral in the usual way by substituting in the Fourier series expansion for $\Delta\phi$ and $\Delta\phi^\ast$
\begin{gather}
\Delta\phi = \frac{1}{L^4}\sum_{n}a_ne^{-ik_{n,\lambda}x^{\lambda}},\\
\Delta\phi^\ast = \frac{1}{L^4}\sum_{n}a^\ast_n e^{ik_{n,\lambda}x^{\lambda}}.
\end{gather}
Evaluating the Gaussian integrals and converting the remaining sum into an integral in the usual way leads to
\begin{equation}
\Delta T^{\mu}_{\;\;\nu} =  -i\hbar\int \frac{d^2k}{(2\pi)^4}\frac{\displaystyle \left\{2k^{\mu}k_{\nu} - \eta^{\mu}_{\;\;\nu}\left(k^2 - m^2\right)\right\}}{\displaystyle k^{2} - m^2}.
\label{vf}
\end{equation}

The integral in Eq. \eqref{vf} is formally divergent. One can regularize the integral via dimensional regularization using the following standard integrals in $D$ dimensions in the limit that $\varepsilon\rightarrow 0$
\begin{widetext}
\begin{align}
\int d^D k\, \frac{1}{(k^2 - m^2 + i\varepsilon)^a} &= i\pi^{D/2}(-1)^a \frac{\Gamma(a - D/2)}{\Gamma(a)} \frac{1}{m^{2a-D}}, \\
\int d^D k\, \frac{k^2}{(k^2 - m^2 + i\varepsilon)^a} &= i\pi^{D/2}(-1)^{a+1} \frac{\Gamma(a - D/2 - 1)}{2\Gamma(a)} \frac{D}{m^{2a-D-2}},\\
\int d^D k\, \frac{k^\mu k_\nu}{(k^2 - m^2 + i\varepsilon)^a} &= i\pi^{D/2}(-1)^{a+1} \frac{\Gamma(a - D/2 - 1)}{2\Gamma(a)} \frac{\eta^{\mu}_{\;\;\nu}}{m^{2a-D-2}}.
\end{align}
\end{widetext}
In this case $a=1$ and $D=4$. Strictly speaking these integrals only exist for $a>D/2$ but we can apply analytical continuation to these expressions to obtain a closed form answer. However, even in this case these integrals do not converge and, hence, we evaluate them in dimension $D=4-2\delta$ to extract the singularity (an infinite vacuum energy) leaving the finite contribution. Putting these contributions together one obtains
\begin{equation}
\Delta T^{\mu}_{\;\;\nu} = \frac{\hbar c}{16\pi^2} m^4\eta^{\mu}_{\;\;\nu}\bigg\{\frac{1}{\delta} + \frac{1}{8}\left(3 - 2\gamma\right) + \mathcal{O}\left(\delta\right)\bigg\},
\end{equation}
where $\gamma$ is the Euler-Mascheroni constant and the factor of $c$ comes from the $k_0$ (energy) integral. The first term is the singularity which has now been extracted from the expression. The second term is finite. All the higher order terms vanish as $\delta\rightarrow 0$. Thus we obtain for the finite contribution to the vacuum fluctuations as
\begin{equation}
\Delta T^{\mu}_{\;\;\nu} = \frac{\hbar c}{128\pi^2}m^4\left(3 - 2\gamma\right)\eta^{\mu}_{\;\;\nu}.
\end{equation}
Converting back to mass in SI units gives
\begin{equation}
\Delta T^{\mu}_{\;\;\nu} = \frac{c^5}{128\pi^2\hbar^3}\bar{m}^4\left(3 - 2\gamma\right)\eta^{\mu}_{\;\;\nu}.
\label{DT}
\end{equation}

\subsection{Application to the Cosmological Constant}

The connection to the Cosmological Constant can be made by noting that the Stress-Energy Tensor is the source term for gravitational curvature. Including the fluctuation in the Stress-Energy Tensor, Einstein's equations without a Cosmological Constant become \cite{EinE}
\begin{equation}
G_{\mu\nu} = \frac{8\pi G}{c^4} \left(T_{\mu\nu} - \Delta T_{\mu\nu}\right),
\label{ein}
\end{equation}
where $G_{\mu\nu}$ is the Einstein Tensor and $G$ is Newton's gravitational constant. Rearranging we have
\begin{equation}
G_{\mu\nu} + \frac{8\pi G}{c^4}\Delta T_{\mu\nu} = \frac{8\pi G}{c^4}T_{\mu\nu}.
\end{equation}
Substituting in the expression for $\Delta T_{\mu\nu}$ from Eq. \eqref{DT} leads to
\begin{equation}
G_{\mu\nu} + \frac{c G}{16\pi\hbar^3}\bar{m}^4\left(3 - 2\gamma\right)\eta_{\mu\nu} = \frac{8\pi G}{c^4}T_{\mu\nu}.
\end{equation}
Comparing this result with Einstein's equations with a Cosmological Constant,
\begin{equation}
G_{\mu\nu} + \Lambda \eta_{\mu\nu} = \frac{8\pi G}{c^4}T_{\mu\nu},
\end{equation}
one finds
\begin{equation}
\Lambda = \frac{c G}{16\pi\hbar^3}\bar{m}^4\left(3 - 2\gamma\right),
\end{equation}
in units of $m^{-2}$. Note that the interpretation of the $\Delta T_{\mu\nu}$ term is different to that of semi-classical gravity \cite{sclg} where the right hand side of Einstein's equations is given by the expectation value of the Stress-Energy tensor operator
\begin{equation}
G_{\mu\nu} + \Lambda g_{\mu\nu} = \frac{8\pi G}{c^4} \langle \hat{T}_{\mu\nu} \rangle.
\end{equation}
The expectation value of a quantum operator describes the statistical average of the quantum matter distribution. The term $\Delta T_{\mu\nu}$ in Eq. \eqref{ein} is the change in the Stress-Energy tensor as one moves away from the classical path and, hence, describes the variance of the quantum matter distribution.

This expression for the Cosmological Constant displays many features of previously proposed expressions for free field contributions. Notably it is proportional to the mass of the field to the fourth power, which is the expected dimensionality \cite{weinberg, martin}. Clearly any accurate numerical value would need to take into account other types of particle field and the interactions between them but it shows that this approach gives results that are consistent with the current calculations of the Cosmological Constant. 

\section{Summary}

Recent work has shown that that Noether's theorem includes quantum fluctuations if one allows the evolution of the system to deviate from the classical path. Here, we have shown that this formalism extends to field theory. Thus, with only knowledge of the classical action, it is possible to obtain the quantum fluctuations of any conserved quantity directly from Noether's theorem. We found the quantum fluctuations in the stress-energy tensor for a complex scalar field and showed how one can relate them to the Cosmological Constant, obtaining an expression that has the correct dimensional features and is consistent with current results.


\end{document}